RESEARCH ARTICLE

# Dynamics of organizational culture: Individual beliefs vs. social conformity

**Christos Ellinas[1,2¤]\*, Neil Allan[2,3], Anders Johansson[3]**

**1** Engineering Mathematics, University of Bristol, Bristol, United Kingdom, **2** Systemic Consult Ltd, Bradford-on-Avon, United Kingdom, **3** Systems IDC, University of Bristol, Bristol, United Kingdom

¤ Current address: Engineering Mathematics, University of Bristol, Bristol, United Kingdom
\* ce12183@bristol.ac.uk

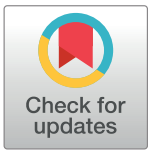

**Data Availability Statement:** All relevant data are within the paper and its Supporting Information files.

**Funding:** CE was partially funded by the Engineering and Physical Sciences Research Council (EPSRC), UK (https://www.epsrc.ac.uk/) under the grant EP/G037353/1, an EPSRC Doctoral Prize fellowship and Systemic Consult Ltd (http://www.systemicconsult.com/). Systemic Consult Ltd provided support in the form of salaries for authors CE and NA, but did not have any additional role in the study design, data collection and analysis,

## Abstract

The complex nature of organizational culture challenges our ability to infer its underlying dynamics from observational studies. Recent computational studies have adopted a distinctly different view, where plausible mechanisms are proposed to describe a wide range of social phenomena, including the onset and evolution of organizational culture. In this spirit, this work introduces an empirically-grounded, agent-based model which relaxes a set of assumptions that describes past work±(a) omittance of an individual's strive for achieving cognitive coherence; (b) limited integration of important contextual factorsÐby utilizing networks of beliefs and incorporating social rank into the dynamics. As a result, we illustrate that: (i) an organization may appear to be increasingly coherent in terms of its organizational culture, yet be composed of individuals with reduced levels of coherence; (ii) the components of social conformityÐpe er-pressure and social rankÐare influential at different aggregation levels.

## Introduction

On July 13[th] 2012, JP Morgan announced a loss of 5.8 billion USD as a result of fraudulent activity taking place, ironically, in a unit aimed in reducing risk [1]. In the following years this event has come to be known as the ªLondon Whaleº incident, with substantial financial consequencesÐincluding 459 million USD of losses in net income and over 1 billion USD in penalties imposed by regulators [1,2], with JP Morgan putting aside a further 23 billion to pay for related potential legal bills to come [3]. One would expect that a single manifestation of such malpractice would have been enough to tarnish the reputation of the entire sector, yet nothing much has changed for the finance sector, with examples of fraudulent activity and misconduct continuing to emerge on a regular basis [4].

The emergence of these events is partly attributed to the *way* in which risk management is practiced, which reflects the risk culture of a given organization. In this context, risk culture can be defined as ªpatterns of behavior, habits of thinking, traditions and rituals, shared values and shared languageº [5] within these organizations. In other words, risk culture can be interpreted as the embodiment of various beliefs that affect the way the risk management function is performed within a given organization. More generally, any set of beliefs that affects a given organizational function—such as risk management—reflects organizational culture.

Understanding organizational culture falls in the class of problems traditionally tackled by social scientists. Such problems are notoriously hard to tackle (i.e. non-linear in nature [6]; multiple levels involved [7]; temporal character [8]), with traditional approaches being limited in identifying correlations between certain variables, whilst noting the ability of certain control variables in mitigating this behavior. However, recent arguments have challenged the validity of such regression studies due to the complex nature of the underlying dynamics, emphasizing the fact that little attention has been given in mapping the underlying *mechanisms* responsible for the emergence of these correlations [6,9,10]. A compounding factor to this criticism is the static view imposed on such social phenomena, with the majority of sociological studies being limited in describing *snapshots* of an organization's state rather than focusing on the dynamic nature of the problem [8,11] (some notable exceptions can be found in the recent review of [12]). The issue with such an approach is that volatile micro level dynamics may be missed by looking at a macro level trend, which in turn may lead to a misleading interpretation of the nature of the system being studied. As a result, social scientists have so far been unable to provide a unified theory for explaining the emergence of collective social phenomena that define organizational behavior [13±15].

Recent developments made under the umbrella of complexity science [16±18] have adopted a distinctly different view, where contextual differences of various social phenomena are abstracted away in search for overarching principles [19]. Such studies typically introduce plausible mechanisms which are subsequently tested in their capacity to replicate widely observed patterns, with homophily and social influence often cited examples of such mechanisms [20,21].

Despite the appeal of such generalization, context dependent aspects are often crucial in the dynamics of collective social phenomena, questioning the extent of abstraction a model should have [22] (for a network-related discussion, see [23]). Hence the challenge lies in identifying mechanisms that capture important aspects of a phenomenon while preserving the model's transparency. In the case of collective social behavior, typical mechanistic models focus on social [24] *or* cognitive [25] aspects, with a choice between the two being enforced in an attempt to keep the proposed model as simple as possible (and consequently preserve generalizability of results). In the context of adopting a new cultural belief, the majority of work focuses either on the process of adopting a new belief due to peer-pressure (the social aspect) or due to increase cognitive coherence (the cognitive aspect). Yet we argue that by decoupling the two aspects, the conflicting reality of certain social phenomena is omitted e.g. an individual may adopt a certain belief due to social conformity even if it contradicts his/her own beliefs. Therefore, this study introduces an integrative framework able to capture both social and cognitive aspects in a simple model. In addition, the model extends the degree of contextual integration by introducing both peer-pressure and social rank (i.e. social conformity) into the core dynamics, thus extending previous studies which focused in examining each aspect in isolation (e.g. [26] and [27] respectively). Notable examples which adopt a similar integrative approach by accounting for both social and cognitive aspects include the recent work of Gavetti and Warglien [28] and Rodriguez et al. [29].

The contribution of this study lies at both a theoretical and practical level. From a theoretical point of view, the development of a formal model of the dynamics of organizational culture can allow for an explicit test of various hypotheses found in the large body of empirical work that has already been developed. In doing so, it has the potential of exposing weaknesses of prevailing wisdom (e.g. individual behavioral traits are independent, as assumed in [30];

collective behavior can be modelled in a context independent manner i.e. ignoring the influence of social rank, as assumed in the class of threshold models [24]) and thus, sharpen future research questions. From a practical point of view, the model can be used to assess how various organizational changes, such as the underlying hierarchical structure, can affect the evolution of organizational culture.

## Literature review

Left alone, it is reasonable to assume that every individual would possess a unique set of beliefs, negating the very notion of shared beliefsÐand  to an extentÐorganizational  culture. Yet it is common experience that beliefs are exchanged between individuals through *social interaction* [17], with individuals reacting accordingly by adopting, amending and/or discarding various beliefs [31±33]. As a result of these actions, the onset of organizational culture can follow a number of possible trajectories, including complete agreement (i.e. every individual shares the exact same beliefs), complete disagreement (i.e. every individual holds a different belief) and various meta-states (i.e. clusters of agreement of various size). In other words, even though there is an envisioned culture at which an organization abides to, achieving coherence at lower aggregation levels (e.g. individuals) is increasingly challenging due to its emergent nature (e.g. [34]).

In an attempt to explore the role of social interaction (or *peer-pressure*) in collective behavior, Granovetter [24] highlighted how individuals are willing to switch behavior, if a given percentage of individual surrounding them already shares that behavior [35]. In other words, the state of an individual is a function of the state of its neighborsÐand  in general, to the social networkÐwith  a threshold value controlling the individual's tolerance to the induced peer-pressure. This powerful notion gave rise to the major class of threshold models [26,36] which has subsequently been used to study a wide range of collective social behaviors [35,37]. By doing so, the influence of the social network architecture has slowly consolidated within the field [38,39], shifting the focus in uncovering the mechanisms that take place across these networks.

The opinion vector-based model is one such class of models that focuses on these mechanisms [17]. In general, opinion vector-based models define the state of an individual as a vector of independent behavioral traits which can be modified through social interactions. A prominent example has been developed by Axelrod [30], which eloquently proposes that: the probability of two individuals interacting is a function of their belief overlap (i.e. *homophily*), with interacting individuals becoming increasingly similar through imitation (i.e. social influence). Despite the evident self-reinforcing nature of Axelrod's model, complete agreement between individual agents is not always attainable, with disparate clusters of distinctly different sub-cultures emerging. The simplicity and non-trivial behavior of this, and similar, models have made it increasingly attractive, with recent work applying it to progressively more realistic contexts, where individuals are embedded in complex network architectures that resemble real-life interaction networks e.g. [32,40,41]. This class of models illustrates the non-trivial outcome of even simple, plausible dynamics that may describe aspects of social behavior, reinforcing the proposition of computational models as suitable tools for exploring organizational behavior [13,42,43]. More generally, it is an early response to recent calls from organization theorists, proposing a shift of focus from mapping the state of an organizational aspect to understanding the dynamics that fuel it.

In summary, threshold models [24,26] and opinion vector-based models [30,32] are two major classes of models that have been used to explain the emergence of empirically-noted correlations across various collective social behaviors. However, a set of assumptions that underlies these models challenge their validity. In particular, opinion vector-based models assume that behavioral beliefs held by an individual are entirely *independent*. This is because the process of trait exchange is modelled on each individual trait independently of the state of the remaining traits. Yet, psychological research has consistently shown that individuals strive for cognitive coherence using various cognitive mechanisms [25,44], suggesting that the converse is true i.e. beliefs are interacting. Hence, by assuming that beliefs are independent, the conflicting nature between preserving internal consistency (by rejecting an inconsistent belief) and peer-pressure (by accepting an inconsistent belief) is missed. Similarly, threshold models consider the structure of the social network as the only factor relevant to the dynamics e.g. [24,35,36]. Hence, these models are context agnostic i.e. contextual information is assumed to be irrelevant to the dynamics of the social interaction process—a typical feature of studies that draw from the natural sciences [22,39]. Yet recent studies challenge the validity of this assumption in an organization context, where the perception of rank plays a key role in collective functions, including organizational learning [27], social exchange [45] and co-operation [46]. In other words, recent work suggests that the strength of social conformity is a function of both social interaction *and* social rank, yet the latter is ignored by the class of threshold models.

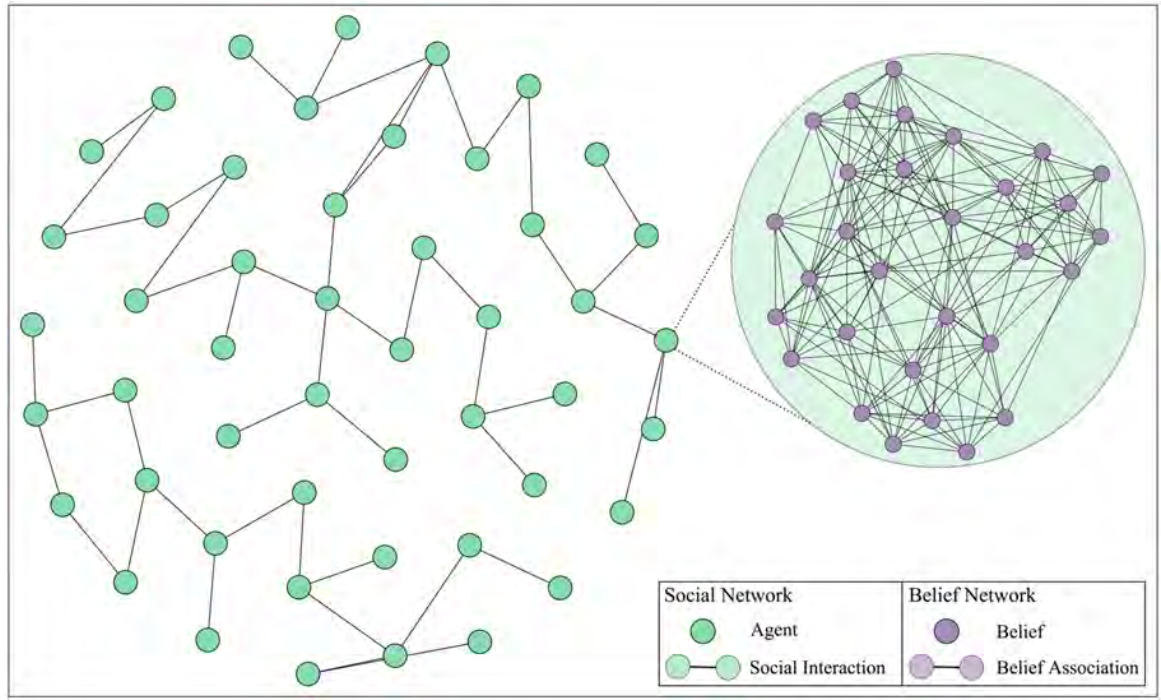

**Fig 1. Social interactions between individuals (green) explicitly capture social effects (e.g. peer-pressure), with each the cognitive state of each individual being captured by a belief network (purple).**

In response, this work develops an integrative, agent-based model where a network of interactions between agents is constructed, with the cognitive state of each agent being characterized by a set of interconnected beliefs—for a visual overview of the model see Fig 1. By doing so, this model relaxes the two aforementioned assumptions (i.e. belief independence and being context agnostic) by: (a) accounting for the conflict between external (peer-pressure) and internal (preserving cognitive coherence) pressure in accepting an external belief, and (b) including both peer-pressure *and* social rank directly into the dynamics of the belief adoption mechanism.

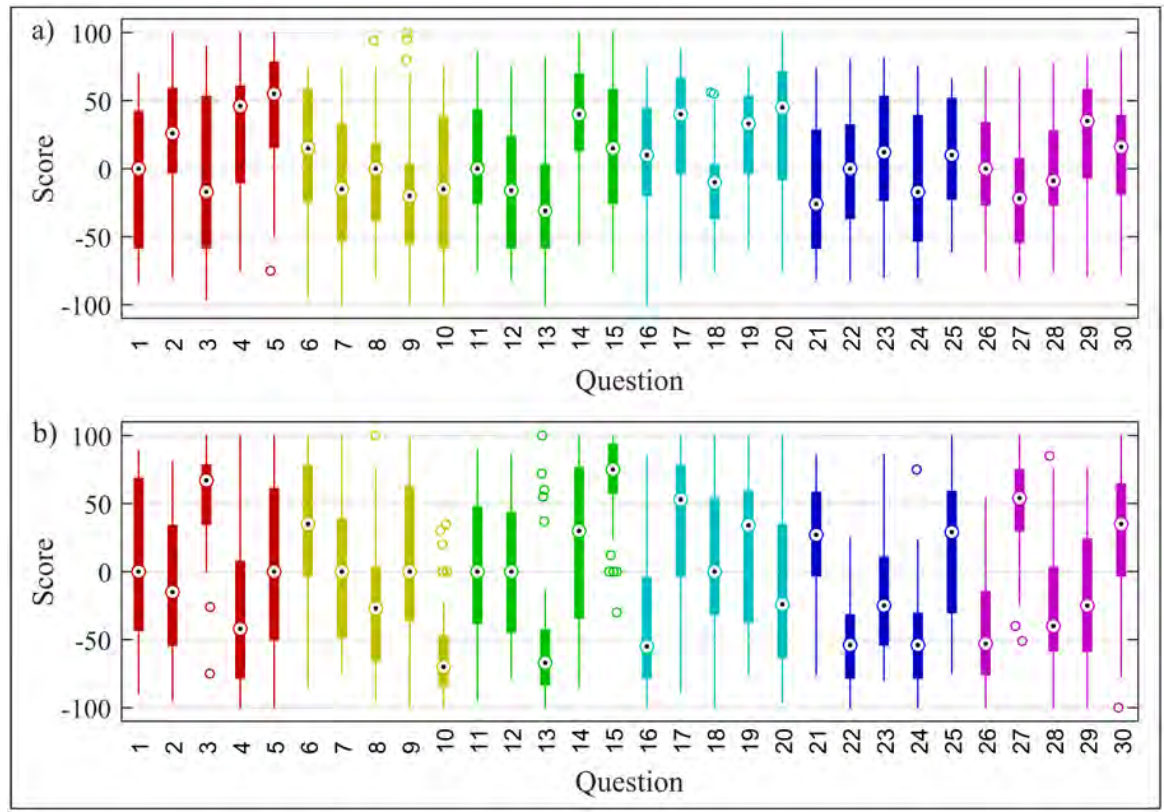

**Fig 2. (a) and (b) reflect response scores for each question for current and desirable (future) state respectively.** Color coding reflects the theme of the question. Central mark in each box plot, with top and bottom box edges, correspond to the median, the 25th and the 75th percentile respectively. Markers outside the box correspond to outliers.

https://doi.org/10.1371/journal.pone.0180193.g002

## Methods

### Belief network

An empirical dataset is used as the basis to construct the belief network of each agent. Specifically, the dataset is composed of survey results captured during a risk-culture mapping project commissioned by a UK-based insurance organization. Each of the 49 participants was given a total of thirty questions revolving around six central themes (five questions per theme), with each question drawing on a specific *belief* related to the application of risk management processes within their organization. For details see S1 File; the entire dataset is available in S1 Dataset. With beliefs beings widely-considered to be a core component of culture [12,47], this dataset can be viewed as a suitable proxy for the risk culture of this organization.

Each question has two components, where each participant is asked to reflect on both current *and* desirable state of that given belief—see Fig 2a and 2b respectively. By doing so, the study captures whether a given individual prescribes ªmore of the sameº behavior—i.e. future state for a belief scores equally, or higher, that its current counterpart—or a shift in the current behavior, referred to as ªless of the sameº i.e. future state for a given belief scores lower than the current state.

In order to relax the assumption of belief independence, the construct of Social Knowledge Structure [48] is used, where associations between beliefs are introduced resulting to a *belief network*. Association assignments follow the structure of a random network with a modular structure reflecting the nature of the survey (i.e. six modules are enforced, corresponding to the number of themes)±see Table A in S1 File. In doing so, every network is inherently composed of stable triads, suggesting that all agents are initially characterized by perfect cognitive. This initialization stage is done to ensure that all agents enter the simulation at the exact same stage, ensuring consistency across all model realizations. Finally, it should be noted that the enforced structure is clearly an assumption, as we are unable to infer the actual belief network from the survey in an unbiased manner. As such, the influence of the belief network topology itself is important and worthy of further exploration.

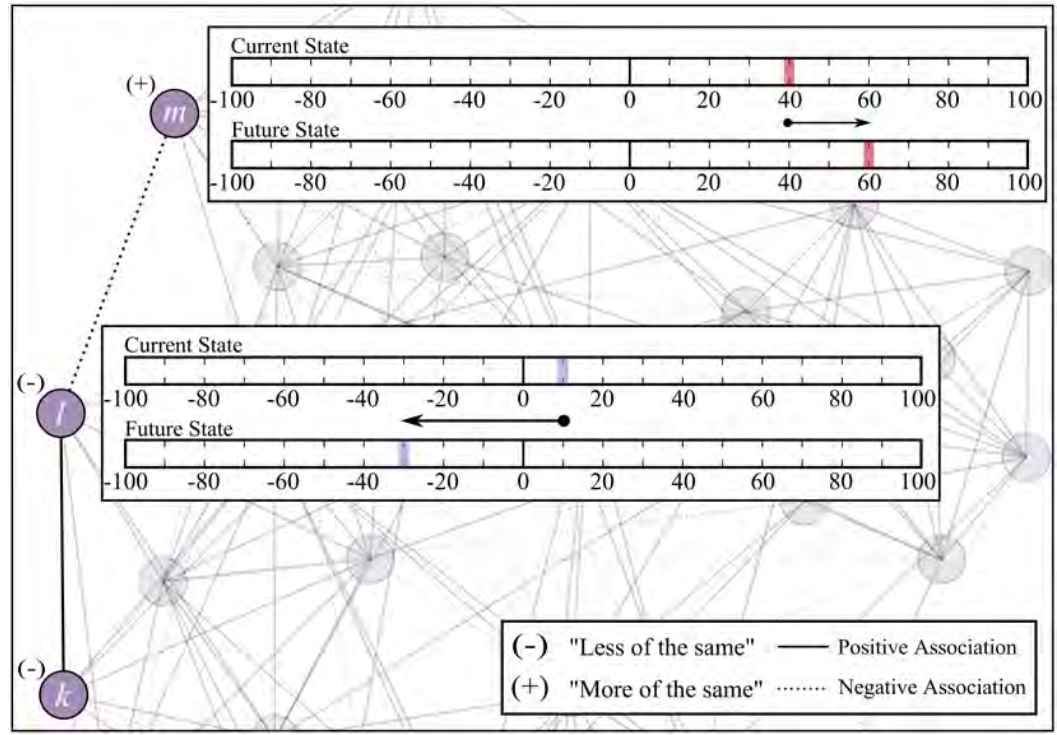

**Fig 3. Belief network for a given individual, where each node corresponds to a belief and links corresponds to belief associations.** Boxes corresponds to typical survey scores, with upper box leading to belief $m$ being allocated a ᵃ+° (ᵃmore of the sameº) while belief $l$ is allocated a ᵃ-° (ᵃless of the sameº). Consequently, an association between similarly signed beliefs (e.g. belief $k$ and $l$) is considered as positive (solid line), with opposite signs (e.g. belief $l$ and $m$) resulting to a negative associations (dotted line).

The survey results are subsequently introduced into the belief network in the following manner: a belief which corresponds to ᵃmore of the sameº behavior is characterized by a positive signÐsee Fig 3, belief $m$; the same applies for the case where a belief has the same current and future state Conversely, a response of ᵃless of the sameº belief, results to belief $l$ being allocated a negative sign. Once each node receives a sign depending on the nature of the belief, every association is signed depending on the nature of the two beliefs that it relates: if the two beliefs have similar signs (i.e. +/+ or -/-), a positive association is obtained (Fig 3; belief $k$ and $l$); in the case of dissimilar signs (i.e. +/- or -/+) a negative association is obtained (Fig 3; belief $l$ and $m$). This process is repeated for each agent at every realization of the dynamics, and in effect results to each agent having a distinct belief network, with respect to all other agents, *and* with respect to itself at different realizations.

Formally, the belief network is defined as an undirected graph $G^{BN} = \{\{V^{BN}\}\ \{E^{BN}\}\}$, where each belief $l$ is abstracted as node $l$, $l \in V^{BN}$. Similarly, an association relating belief $l$ and $m$ is represented by an undirected, signed link $e_{b,m}$ where $\{e_{l,m} \in \{-1, 1\} | e_{l,m} \in E^{BN}\}$. To specify the participant owing a specific belief network, a superscript is used; consequently belief $l^i$ and association $e^i_{l,m}$ refer to the belief network of participant $i$. Note that for every realization of the model, each agent is attributed a new belief network.

## Social network

To account for the role of social conformity in the process of belief exchange, individuals are embedded in a network structure which represents the social interactions between them. Formally, the social network is defined as an undirected graph $G^{SN} = \{\{V^{SN}\}\{E^{SN}\}\}$ where each participant $i$ of the survey is abstracted as agent $i$, $\in V^{SN}$, with every interaction between agent $i$ and $j$ being represented by the undirected link $e_{i,j}$ where $e_{i,j} \in E^{SN}$.

A generative model able to replicate typical characteristics of social networks is used. Specifically, empirical studies have highlighted the importance of clustering in social networks where the effect of collective dynamics is prominent [24,49,50]. Seminal work by Watts and Strogatz [51] has further consolidated this insight by identifying it across a wide range of networks, with further work illustrating the perseverance of such architectures across various relevant domains [52]. Watts and Strogatz [51] further introduced a generative model capable of replicating the effect, which is subsequently used to generate the social network used herein. Specifically, the generative process is grounded on two basic steps: (a) construct a lattice—in a ring formation—with a given number of connections, and (b) randomly rewire a given portion of link in order to introduce ªshortcutsº between distant nodes.

In order to generate this network, a third parameter is needed, which corresponds to the average degree of each node. This parameter is effectively used to set aspect (a)±in this case it is set to 2 i.e. each individual regularly interacts directly with a further two individual, or roughly 4% of the total organization. With respect to (b), the probability of rewiring a link between two nodes is set to 0.5. Finally, the number of nodes is fixed to reflect the number of survey participants (i.e. 49), with a new social network being generated for every realization of the model, in step with Section 3.2.

## Dynamics

The rules dictating the dynamics of the agent-based model are as follows: at each time step t, a random pair of connected agents $i$ and $j$ is chosen, with agent $i$ (source) randomly choosing an association from its internal belief network and sending it agent $j$ (receiver). Assuming the receiver is willing to listen to the source, the receiver will accept the incoming association if it increases the coherence of its belief network. If not, the receiver may still accept the incoming association based on social grounds i.e. the individual foregoes cognitive consistency for social conformity. In the case where the incoming association is accepted, it may have one of the following effect—it serves as a new association between two existing beliefs *or* it replaces an old one association. The probability for accepting an incoming association is a function of peer-pressure (quantified as the portion of the receiver's neighbors that agree with that belief) and social rank difference between the receiver and its neighbors. A control parameter $\gamma$ is introduced to control the influence balance between peer-pressure and social rank, enabling various organizational contexts to be formulated and subsequently tested (see Eq 3).

Formally, the component of the probability function responsible for introducing peer-pressure in the mechanism is defined as:

$$P^{PP} = \frac{1}{|B(j)|} \Sigma_{k \in B(j)} \frac{1}{|\Gamma(k)|} \Sigma_{l,m \in \Gamma(k)} \delta^k_{l,m}, \text{ where } \delta^k_{l,m} = \begin{cases} 1, & \text{if } e^j_{l,m} = e^k_{l,m} \\ 0 & \text{otherwise} \end{cases} \quad (1)$$

where the sum is taken over all the neighbors of agent $j$ (denoted by set $B(j)$), running across the set of associations that characterize the belief network of every agent found in $B(j)$Đthis set of associations is denoted as $\Gamma(k)$. The Kronecker delta $\delta_{l,m}^k$ is used to enumerate the number of associations present in the belief network of agent $k$ that resemble the association proposed to agent $j$, taking a value of 1 only if the two have the same sign (in effect, being identical). Note that this is obtained in a fraction form by dividing over the total number of associations found in the belief network of every agent in $\Gamma(k)$.

Similarly, the component of the probability function that introduces the influence of social rank is defined as:

$$P^{SR} = 1 - \exp\{-F^{SR}\},$$

where

$$F^{SR} = \frac{1}{|B(j)|} \Sigma_{i \in B(j)} \max\{p^i - p^j, 0\} \qquad (2)$$

where the $p^i$ is the social rank of agent $i$, inferred from the survey demographics (see Figure A in S1 File) while the max function ensures that no contribution takes place if the social rank of agent $j$ is higher than agent $i$. The functional form of $F^{SR}$ encapsulates the concept of ªpower-distanceº, as argued within the organizational studies [53,54]. The functional form of $P^{SR}$ has a number of desirable features (e.g. $P^{SR} \in [0,1]$; saturates fast, reflecting the ªtype of power [which] involves someone getting another person to something that he or she would have not otherwise done. They are simply told what to do `orelse'º [55] which is often observed in organizations [56].

To account for both aspects (peer-pressure; social rank), the probability for agent $j$ (receiver) to accept a belief association proposed by agent $i$ (source) is given by:

$$P(e_{l,m}^j = e_{l,m}^i) = \gamma P^{PP} + (1 - \gamma) P^{SR} \qquad (3)$$

where $\gamma$ serves as the aforementioned control variable. As such, the mechanism is driven solely by peer-pressure when $\gamma = 1$, while social-rank is the sole relevant aspect when $\gamma = 0$. For $0 < \gamma < 1$ the Eq 3 allows a mixture of the two components, where the relative strength is controlled by $\gamma$.

The state of the system at any point in time can be characterized at: (a) an agent-level and (b) network-level. With a focus on (a) the principle of triad stability [57,58] can be used to define, and subsequently assess, the cognitive consistency of every agent (triad stability can be interpreted as follows [59]: (i) my friend's friend is my friend; (ii) my friend's enemy is my enemy; (iii) my enemy's friend is my enemy and (iv) my enemy's enemy is my friend.. In this context, the existence of unstable triads increases the psychological discomfort of each agent. This effect can be subsequently alleviated if the agent is given the opportunity to implement some form of dissonance-reduction strategies [60]±in this case by ignoring the incoming belief. The effect of such strategies can be replicated by reducing the number of unstable triads in the agent's belief network and can be achieved by accepting a belief from a neighboring agent. Formally, the cognitive consistency of agent's $j$ belief network ($C_j^{BN}$) is defined as:

$$C_j^{BN} = \frac{1}{N_\Delta} \Sigma_{k,l,m}(e_{k,l}e_{l,m}e_{m,k})\delta_{k,l,m}, \, C_j \in [0, 1] \qquad (4)$$

$$\delta_{k,l,m} = \begin{cases} 1, & \text{if } e_{k,l}e_{l,m}e_{m,k} = 1 \\ 0, & \text{if } e_{k,l}e_{l,m}e_{m,k} = -1 \end{cases}$$

where $N_\Delta$ is the total number of triads in the belief network, $e_{k,l}$ is the signed link between belief k and l (and so on. . .), with $\delta_{k,l,m}$ being the Kronecker delta, filtering stable from unstable triads. This quantity can be used to construct a global measure by averaging $C_j^{BN}$ across all agents, denoted as $\langle C^{SN} \rangle$.

The consistency between pairs of agents, in terms of their belief network, can be used to characterize the homogeneity of the entire social network. To do so, the absolute relative difference between two connected agents, in terms of stable triads (and hence, cognitive consistency) can be used to assess the homogeneity across the neighborhood of agent $j$ ($C_j^{SN}$). Under this definition, a lower $C_j^{SN}$ value suggests increased similarity between a pair of connected agents, in terms of cognitive dissonance:

$$C_j^{SN} = 1 - \frac{1}{|B(j)|} \Sigma_{i \in B(j)} \frac{|C_i - C_j|}{\max\{C_i, C_j\}}, \ C_j^{SN} \in [0, 1] \tag{5}$$

Finally, $C_j^{SN}$ can be used to construct a global measure of network coherence in terms of cognitive coherence by averaging $C_j^{SN}$ it across all agents, denoted as $\langle C^{BN} \rangle$.

## Results

### Conflicting dynamics

The model proposed is characterized by conflicting dynamics, where agents strive for cognitive consistency yet may forego it for the sake of social conformityÐthe latter being a twofold aspect combining elements of peer-pressure and social rank. Additionally, control variable γ effectively dictates the balance between the two (Eq 3). Fig 4 first presents results at the three intermediate states of the model, with 4a, 4b and 4c capturing the network coherence ($C^{SN}$) and average cognitive coherence ($C^{BN}$) at γ = 0, γ = 0.5 and γ = 1.0 respectively. Note that the colored band around each line plot maps the standard error across the independent runs, calculated as the sample standard deviation divided by the square root of the number of runs.

With respect to the case where social rank sole determining factor for adopting an incoherent belief Ðsee Fig 4a. In particular, both network coherence and average cognitive coherence exhibit increasingly similar behaviorÐsee Fig 4a. In particular, both measures follow a monotonically decreasing trend, with network coherence decreasing at a faster rate. After roughly 45,000 time steps, both measures stabilize, reaching values of approximately 0.798 and 0.686 for the average cognitive coherence and network coherence respectively. Interestingly, the fact that cognitive coherence of the average is consistently preserved at higher levels compared to network coherence is an example of how social interactions can undermine the state of the overall system. In the context of risk culture, the case of social rank being the sole determining factor results in a situation where fairly coherent individuals (with respect to their belief system) interact to give rise to an increasingly heterogeneous organizationÐan increasingly undesirable state.

Shifting focus to the case where social rank and peer-pressure play an equal role (i.e. γ = 0.5), both network coherence and average cognitive coherence initially exhibit a monotonically decreasing behaviorÐsee Fig 4b. However, given enough time this convergence breaks down, with the average cognitive coherence of the agents continues to reduce until it stabilizes around 0.63. At the same time, the trajectory of the network coherence of the agents is reversed, exhibiting a slow but steady increase reaching a maximum value of just below 0.9. By the end of the simulation, the majority of connected agents are, on average, increasingly similar to their neighbors (hence, high network coherence) yet each individual agent is increasingly incoherent in terms of its belief network (hence, low average cognitive coherence). In the context of risk culture, the inclusion of both social rank and peer-pressure, at equal weights, results in a situation where the organization appears to be in an increasingly coherent state, despite its composition of increasingly dissimilar individuals.

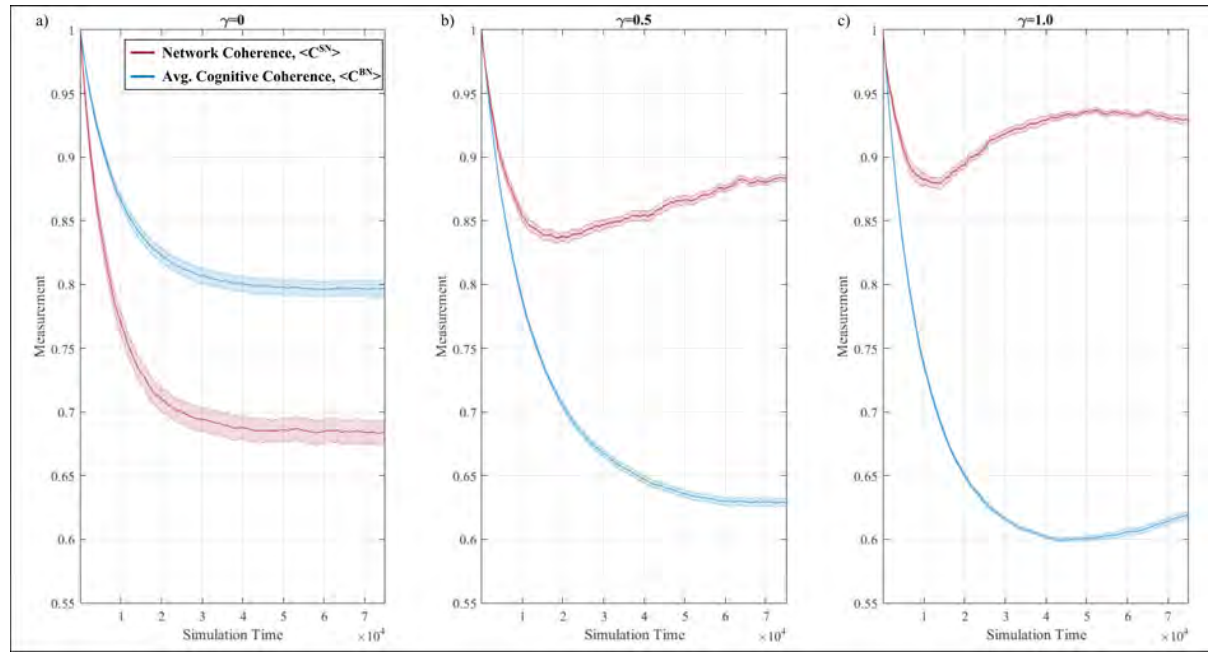

**Fig 4. Evolution of average cognitive coherence (blue) and network coherence (red) across time (x-axis) during the belief exchange process at (a) γ = 0 (social rank is the sole determining factor), (b) γ = 0.5 (social rank and peer-pressure have equal weighting) and (c) γ = 1 (peer-pressure is the sole determining factor).** Note that values begin at 1 due to perfect initial coherence across agents. Band around each plot corresponds to the standard error across twelve independent runs.

Finally, in the case where peer-pressure is the sole determining factor for adopting an incoherent belief (i.e. γ = 1), the disparity between average cognitive coherence and network coherence increases, with the overall behavior becoming non-monotonic—see Fig 4c. In particular, both average cognitive coherence and network coherence initially exhibit a monotonic decrease, albeit at a faster rate compared to the case of γ = 0.5. Cognitive coherence continues to decrease until reaching a minimum value of roughly 0.6 ±at this point the behavior reverses reaching a final value of 0.62. Similarly, the network coherence reverses its decreasing trajectory early on, yet this increase manifest at an increasingly smaller rate until it plateaus at roughly the same time when cognitive coherence reaches a value of 0.6. Beyond this point, network coherence starts to slowly decline, reaching a value of roughly 0.93. By the end of the simulation, the situation is fairly similar to the one obtained in the case of γ = 0.5, where the network appears to be increasingly coherent, despite the overall reduction in cognitive coherence of the individual agents. Importantly, this case varies from the case of γ = 0.5 in the way both measures evolve across time, where Fig 4c suggests a mismatch between the evolution time of cognitive coherence and that of network coherence, evident by the difference in the rate at which the trend of each measure changes. For the sake of completeness, Figure B in S1 File contains additional results for the entire range of γ.

## Organizational vs. individual level

The distinct focus of the two measures introduced herein—average cognitive coherence and network coherence—allows for a systematic examination of the influence of γ across the organizational and individual level. Specifically, Fig 5a illustrates the influence of the belief exchange process at the organizational level by considering network coherence. On the other hand, Fig 5b focuses on the individual level by considering the average cognitive coherence. As such, it is clear that varying γ has a distinctly different effect in terms of aggregation levels, where an increase in γ has a positive effect at the organizational level (denoted by an increase in network coherence) while γ has a negative effect at the individual level (denoted by an increase in average cognitive coherence).

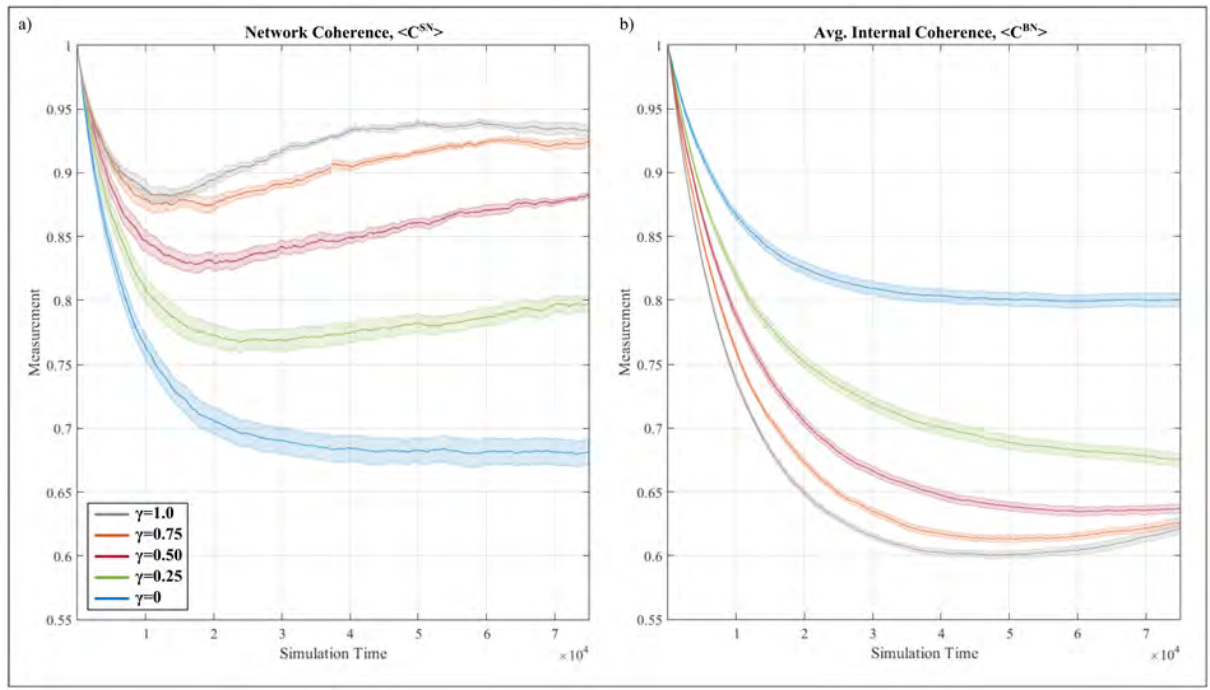

**Fig 5. Impact of varying γ at the (a) network and (b) individual level, using network coherence and average cognitive coherence respectively.** Similar to Fig 4, values begin at 1 due to perfect initial coherence across agents.

Delving further into each aggregation level, the influence of γ impacts both the overall trend and the resulting value. Focusing on the network level (Fig 5a), the monotonic pattern noted in the case of γ = 0 quickly changes to a non-monotonic trend, where the rate in which network coherence increases depends on γ. With respect to the individual level (Fig 5b), a similar change is noted albeit at higher γ values where the overall behavior switches from monotonic to non-monotonic. Considering the distinct effect that γ has with respect to the affected scales (positive effect on the overall network; negative effect on the average individual), the magnitude of difference across its extreme values is also examined. In order to capture this effect, the absolute difference between the two extreme values (i.e. γ = 0 and γ = 1) for each measure is plotted—see Fig 6. Overall, the impact of γ in terms of absolute size is initially greater at the network level (i.e. red line overcomes the blue line). After roughly 20,000 time steps, this behavior changes as the impact of γ at the individual level increases (i.e. blue line overcomes the red line).

The overlap between average cognitive coherence and network coherence is a quantity of particular interest as it indicates whether disparity on the effect of cultural exchange across aggregation levels exists. In other words, if the difference between the trajectories of the two measures is low, it suggests that one may infer the heterogeneity of an organization, in terms of its risk culture, by simply observing its individual. However, if the difference between the two measures is great, such inference breaks down due to way individuals interact. The

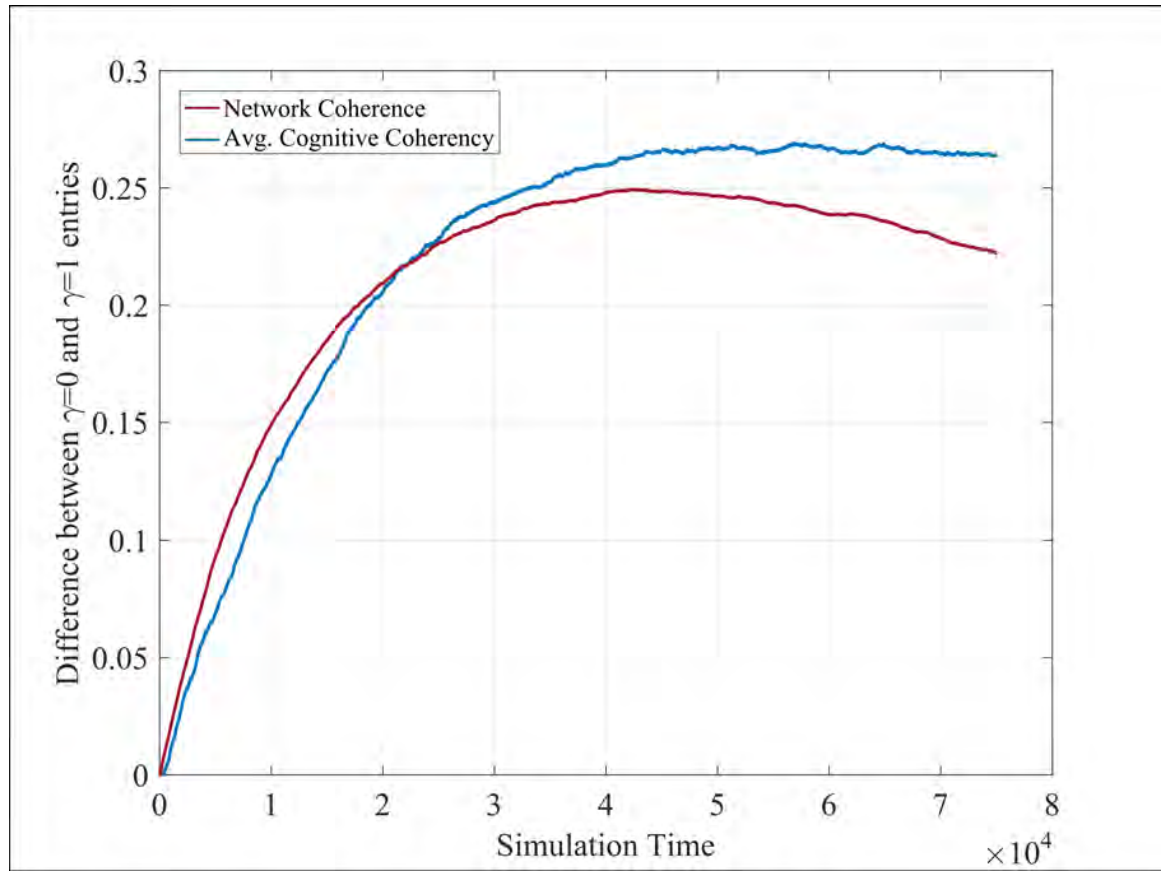

**Fig 6. Quantifying the magnitude of varying γ by capturing the relative difference between tis two extreme manifestation (i.e. γ = 0 and γ = 1) at two levels–network (red) and individual (blue) level.**

difference between the two measures can be captured by using:

$$E = \frac{\langle C^{BN} \rangle - \langle C^{SN} \rangle}{\langle C^{BN} \rangle + \langle C^{SN} \rangle} \ , \ E \in [-1, 1] \tag{6}$$

where $E \in \mathbb{R}^+$ indicates that the average cognitive coherence is higher than network coherence, with $E \in \mathbb{R}^-$ capturing the converse–both cases indicate increased disparity between the individual and network level. A value of $E \approx 0$ indicates that the two levels converge and hence, inference of the state of the organizations from its individual agents.

Fig 7 plots E for the entire range of γ, where solid and dotted weights represent positive and negative E values respectively. With respect to the one extreme (γ = 0), E increases slowly as the simulation progresses, which translates to social average cognitive coherence being larger than the network coherence measure. This difference stabilizes at the final stages of the simulation, reaching a value of approximately 0.07. At this point, the state of an average individual can be inferred by observing the state of the entire organization, as the two measures (average individual coherence and network coherence) are fairly close to each other. However, this situation changes rapidly with increasing γ. In the case of the other extreme end (γ = 1) the value of E dives into the negative regime reaching a maximum value of approximately -0.22, effectively translating to the converse effect i.e. network coherence is higher than average cognitive coherence. At this point, an organization may appear to be increasingly coherent (i.e. high average network coherence)±a rather deceiving deduction since it is composed by individuals with reduced levels of cognitive coherence. More generally, given enough time for the evolution process to set in, it is increasingly challenging to infer the coherence of individuals by mapping the overall organization (and vice-versa). Increasing the role of peer-pressure (i.e. γ increases) amplifies this effect as it results to an increase rate of change for E.

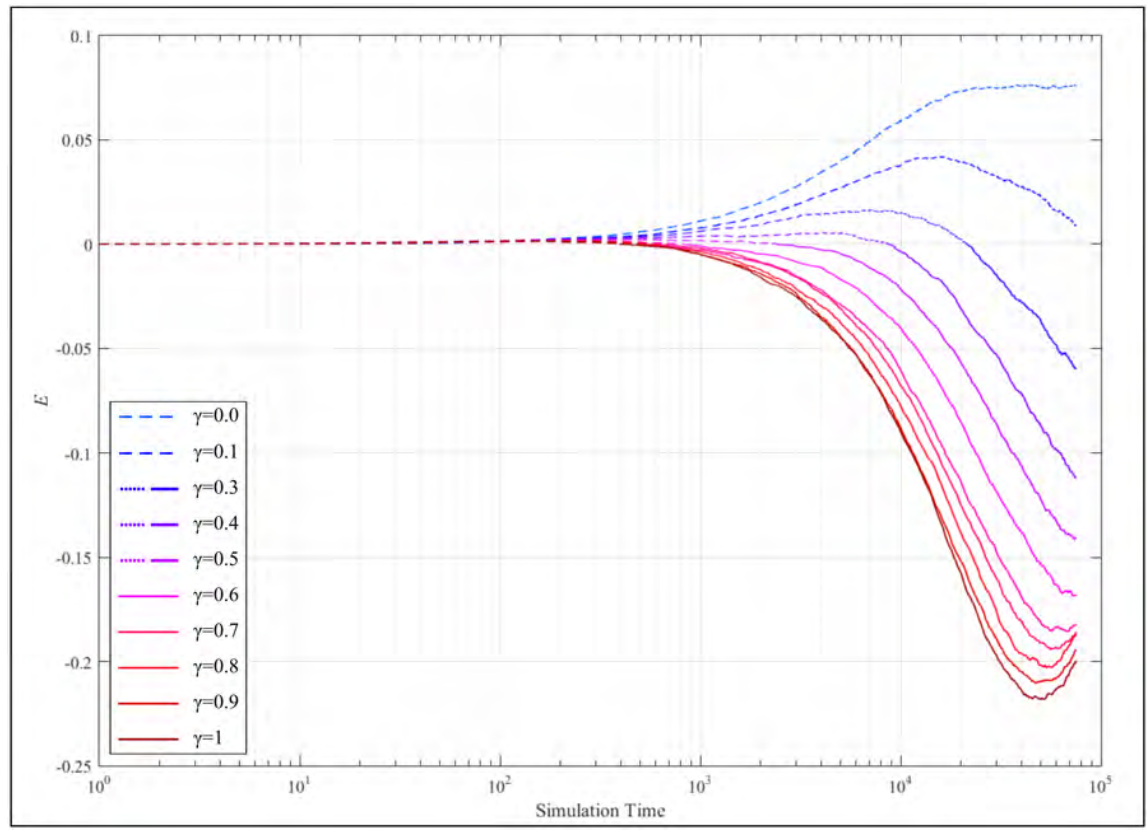

**Fig 7. Trajectory of E at different values of γ, where solid and dotted weights represent positive and negative E values respectively.**

## Discussion

Organizational culture emerges via the aggregation of beliefs of each agent that the organization is composed of. Driven by social interactions and cognitive biases, agents exchange, amend and/or discard their beliefs and as a result, organizational culture remains in a state of continuous flux.

From a methodological standpoint, the proposed model relaxes a number of assumptions that underlie a significant portion of past related work, including belief independence (as assumed in opinion, vector-based models e.g. [30,32]) and being context agnostic (as assumed in threshold models e.g. [24,26]). In particular, the proposed model leverages the conflicting dynamics of individual consistency and social conformity to map the onset of organizational culture. In doing so, it integrates both social and cognitive aspects of the belief adoption processÐaspects which are traditionally examined in isolation. With a focus on social conformity, the compounding effect of its components (peer-pressure and social rank) is isolated and further explored. Results indicate that peer-pressure plays a significant role in shaping the ability of individuals to reduce cognitive dissonance that describes their beliefs. At the same time, social rank significantly affects the homogeneity of an organization, in terms of overlapping beliefsÐsuch insight is in step with the increased recognition of rank as a key determinant to organizational behavior [27]. As such, the proposed model contextualizes the influence of social influence by introducing social rank, and in doing so highlights the disparity in their influence across different organizational levels.

## Theoretical implications

**Disparity between organizational levels.** At the original model formulation (Eq 3, $\gamma$ = 0.5), peer-pressure and social rank influence an agent from adopting a new association, even if it contradicts its own belief network. As a result, two important features are uncovered: (a) at the initial stages of the simulation, the average individual and network coherence decrease monotonically at a similar rate (Fig 4b; ~1,000 time-steps) and (b) given enough time, the two measures diverge, highlighting a chasm between the state of the organization at the individual and network level (Fig 4b; >1,000 time-steps).

With respect to (a), the heightened influence of social conformity in the process of association exchange is linked with a deteriorating coherence in terms of cultural overlap, both at the individual and organizational level. Such insight is consistent with empirical work highlighting the negative relationship between increased exposure to social conformity (e.g. the effect of open-plan offices) and trust (a component of organizational culture)±see [61]. Whilst one should be cautious when drawing such broad inference, the dynamics proposed herein provide a simple, and plausible, mechanism for such phenomena. Importantly, the state of an average individual can be inferred by observing the state of the entire organization, as the two measures (average individual coherence and network coherence) deteriorate at the same rate.

However, such inference is not always possible. Specifically point (b) highlights that an organization may appear to be increasingly coherent (i.e. high network coherence) yet be composed by individuals with increased cognitive dissonance (i.e. low average cognitive coherence). Such individuals may eventually undertake harmful actions, surprising outside observers who were deceived by the evident coherence of the organization. More generally, given enough time for the evolution process to set in, it will be increasingly challenging to infer the coherence of individuals by mapping the overall organization. As a result, damaging events undertaken by individuals with distinctly different culture (e.g. conduct risk [62]) is inherently hard to predict given that the identification of such individuals must take place at an individual basisÐa resource intensive and challenging task. These finding highlight some of the limitations of observational studies for theory building purposes, as (a) a trend appears to emerge and then disappear without any external intervention (where observational studies are limited in mapping the state of an organization at a given point in timeÐsee Roe [8] and Holme and Liljeros [9]) and (b) different organizational levels exhibit distinct behaviors despite the fact that the exact same mechanism is in place (where observational studies do not explicitly distinguish between multiple levels of analysis, as noted by Kozlowski et al. [7]).

**The role of peer-pressure and social rank.** Results highlight that the influence of peer-pressure and social rank are segregated across aggregation levels, where peer-pressure has a greater influence on the overall network while social rank has a greater influence on the state of the individual.

The theoretical argument emerging from this finding is two-fold. Firstly, the influence of social conformity is not isolated on a single organizational level, highlighting the non-trivial nature of its effect. As such, future studies around social collective behavior in general (and organizational culture in particular), should account for distinct levels of analysisÐan argument echoed by Kozlowski et al. [7] and evident by the results of this work. Secondly, studies focusing solely on the influence of the network structure[26] or on the influence of social rank[27] should not be taken in isolation as their influence is exercised at distinct levels.

This point is increasingly important as organizational studies increasingly embrace the complex nature of organizational and are consequently tempted to decouple the two aspects. Results herein further reinforce this argument by providing a comparable richer picture. Specifically, in the case of isolating the effect of peer-pressure and social rank, strictly monotonic behavior typically describes results at both individual and network level. Yet when their effect is integrated, non-trivial behavior is observed i.e. through non-monotonic trend (e.g. Figs 4 and 6). Evidence of this sort emphasize the non-trivial effect of the evolution of organizational culture even under the relatively simple premise of the proposed model.

## Limitations

This work has some limitations that provide opportunities for further work. One limitation is that it considers static network topologies. In the case of the social network, this is a simplification as individuals come and go in an organization (corresponding to a change in the number of nodes in the social network), along with their interactions dynamically evolving (e.g. [63]); a similar argument applies for the belief network. This assumption does not diminish the value of this work as the emphasis here is on integrating the dynamical processes that drive the evolution of the organizational culture—yet providing an increasingly realistic picture of the network that these dynamics unravel upon adds a desirable layer of realism. Adaptive networks may serve as possible route for relaxing this restriction, where dynamics are coupled with the network topology, resulting in an adaptive topology which evolves over time—the work of [63,64] serve as notable examples of their application.

Another limitation stems from the integrative nature of the model, as it provides for a wider range of parameters that can potentially affect the outcome of the model. Specifically, consider the assumed structure of the two network (i.e. social and belief network)±even though they form reasonable approximations [52], the sensitivity of the dynamics in varying the initial parameters that dictate network characteristics remains unexplored. Future work could explore this aspect by considering various topologies (e.g. random, scale-free, core-periphery etc.) using a range of parameters to explore whether any significant differences emerge.

## Managerial recommendations

Varying the hierarchical structure of an organization forms a reasonably form of intervention in an attempt to promote (or hinder) a given evolutionary trajectory of an organization's risk culture. The influence of such intervention can be explored by varying the social rank distribution across the organization. As an example, consider the case where the distribution of social rank resembles a Normal distribution—in effect it implies that the majority of individuals have the same rank, with few deviating on higher and lower levels of hierarchy. In other words, it resembles a relatively `flat' organization. In contrast, consider the case where social rank is distributed based on a Log-Normal distribution—in effect it suggests that the majority of individuals are found on the lower levels of the organization with a few being on much higher levels. In other words, it resembles an increasingly `vertical' organization.

Generally, all three cases of hierarchical structure illustrate qualitatively similar behavior albeit being quantitatively different. In terms of network coherence, all three cases follow a similar, non-monotonic trend. Overall, the case of Log-Normal hierarchy results in reduced performance under both measures, with the Empirical and the Normal case being increasingly close in terms of both measures. In particular, both Normal and Empirical result in similar network coherence values, with this convergence breaking down at the latter stages of the simulation. At this point, the Normal case reaches the highest network coherence value, followed by the Empirical and the Log-Normal caseÐsee Fig 8a. In terms of the average cognitive coherence, the Log-Normal case results to distinctively lower levels of average cognitive coherence, with both Empirical and Normal cases achieving increasingly similar levels (Fig 8b). As such, an increasingly `flat' hierarchy promotes heightened levels of homogeneity and cognitive coherence in terms of belief exchange, while an increasingly `vertical' organization hampers both cognitive coherence and network coherence, inevitably affecting the dissemination of beliefs.

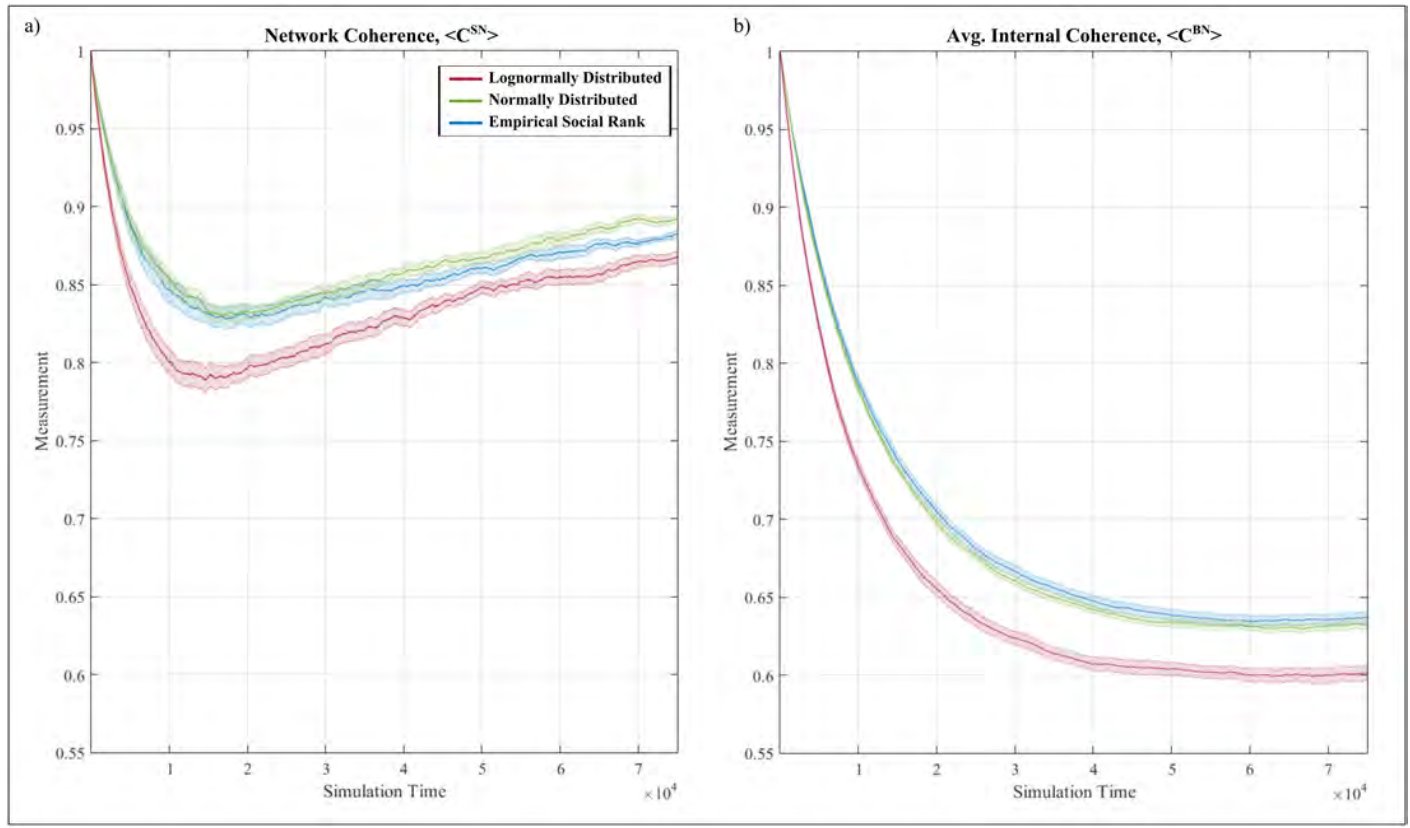

**Fig 8. (a) Network coherence and (b) average cognitive coherence under two distinct hierarchical structuresÐ A `flat' hierarchy (green plot) and a `vertical' hierarchy (yellow plot).** The empirically-obtained hierarchical structure is included for reference (blue line). Results corresponds to the case of γ = 0.5.

It is worth noting that the relevance of these results extends beyond the dissemination of risk-related cultural beliefs to the general dissemination of various quantities across an organization. For example, increased levels of both cognitive and network coherence are bound to increase the rate of information exchange and hence accelerate collective functions such as organizational learning. In such context, an increasingly `vertical' organization hinders both individual coherence and network coherence, consequently hampering collective functions. Such evidence is consistent with recent empirical studies that highlight a negative relationship between increased organizational structure (in the form of hierarchical levels) and internal team learning [65,66], whilst providing a plausible mechanism that may be responsible for noting such effects.

## Conclusion

In this paper, we have proposed an empirically-grounded, integrative model that was used to tackle the following previous assumptions; (a) belief independence; (b) increasingly context agnostic; by utilizing networks of beliefs and incorporating social rank (which is an important aspect in the context of organizations).

Thereby, results indicate that increased social conformity can be increasingly damaging to the evolution of organizational culture—a view consistent with past empirical work [61]. In the context of organizational hierarchy, a `flat' organization outperforms a `vertical' organizational structure in terms of culture coherence, benefiting related processes such as organizational learning. Such insight is consistent with recent empirical work [65,66] reinforcing the plausibility of the proposed model. By isolating the influence of peer-pressure and social rank, a disparity of scales, in terms of their influence, emerges, with peer-pressure having a greater impact on the macro scale (i.e. organization) while social rank has a stronger influence at the micro level (i.e. individual). As a result, future attempts focusing on the influence of social conformity to organizational behavior should follow similarly integrative approaches otherwise they risk missing the interplay of influence between the macro and micro organizational levels.

## Supporting information

**S1 File. Additional supporting information.** Document contains additional clarifications and results. **Table A. Central themes of survey**. Spectrum of the six central themes on which the survey builds on. **Figure A. Power rank histogram of individuals**. Histogram of individuals' power rank, which corresponds to their role and experience within the organization. **Figure B. Additional results for entire range of $\gamma$.** Additional results, with respect with respect to average cognitive coherence and network coherence, for the entire range of $\gamma$, $\gamma \in [0,1]$. (DOCX)

**S1 Dataset. Survey responses.** Dataset containing survey answers used to initialize the model. (XLSX)

## Author Contributions

**Conceptualization:** CE.

**Data curation:** CE.

**Formal analysis:** CE.

**Funding acquisition:** CE NA AJ.

**Investigation:** NA.

**Methodology:** CE.

**Project administration:** CE AJ NA.

**Resources:** CE.

**Software:** CE.

**Supervision:** AJ NA.

**Visualization:** CE.

**Writing ± original draft:** CE.

**Writing ± review & editing:** CE AJ NA.